\documentclass[letterpaper, 10 pt, conference]{ieeeconf}  

\IEEEoverridecommandlockouts                              
\title{\huge \bf
A Survey on Signed Graph Embedding: Methods and Applications
}

\author{ \parbox{3 in}{\centering Shrabani Ghosh 
        \thanks{*Use the $\backslash$thanks command to put information here}\\
        Computer Science \\
        University of North Carolina Charlotte\\
        {\tt\small sghosh15@uncc.edu}}
}

\usepackage[justification=centering]{caption}
\usepackage{graphicx}  
\usepackage{float}
\usepackage{amsmath}
\usepackage{amssymb}
\usepackage{hyperref}
\usepackage{tabularx}
\usepackage{tikz}
\usepackage{comment}
\usepackage{subcaption}
\usepackage[utf8]{inputenc}
\usepackage{lastpage}
\usepackage{adjustbox}

\usepackage{lipsum}
\begin{document}
\pagenumbering{alph}

\maketitle
\thispagestyle{empty}
\pagestyle{empty}

\begin{abstract}

A signed graph (SG) is a graph where edges carry sign information attached to it. The sign of a network can be positive, negative, or neutral. A signed network is ubiquitous in a real-world network like social networks, citation networks, and various technical networks. There are many network embedding models have been proposed and developed for signed networks for both homogeneous and heterogeneous types. SG embedding learns low-dimensional vector representations for nodes of a network, which helps to do many network analysis tasks such as link prediction, node classification, and community detection. In this survey, we perform a comprehensive study of SG embedding methods and applications. We introduce here the basic theories and methods of SGs and survey the current state of the art of signed graph embedding methods. In addition, we explore the applications of different types of SG embedding methods in real-world scenarios. As an application, we have explored the citation network to analyze authorship networks. We also provide source code and datasets to give future direction. 
Lastly, we explore the challenges of SG embedding and forecast various future research directions in this field.


\end{abstract}

 \section{\textbf {\Large INTRODUCTION}}
A graph is a powerful tool that can represent ubiquity of data such as biological, social, and technological systems. In such graphs, the actors are comprised of nodes and their relationship/interactions comprise the edges of the graph. In an unsigned network, all edges denote a positive relationship. The typical unsigned graph, for example, online social networks such as Facebook, where users are nodes and edges represent the bidirectional friendship relationship. Though most of the datasets are modeled as unsigned networks, other information is more important to represent the relationship. In many networks, the interaction between agents can have relationships that can be either positive or negative and these are better represented as a signed network. Signed graphs can have two types of edge signs: positive and negative. Sign graphs are widely seen on social media sites where positive sign reflects support, friendship, and agreement, and negative sign reflects disagreement, and antagonism. For example, Slashdot \cite{leskovec2010signed}: a product review dataset where users mark their trust and distrust to other users, Epinions \cite{leskovec2010signed}: online review social network that users express their views about diverse items like music, hardware, TV shows, Wikipedia \cite{leskovec2010signed}: a voting dataset where user nominated Wikipedia members to elect as administrators. In graph analysis, graph representation learning is a primary research problem for common tasks, e.g., node classification \cite{tang2016node}, link prediction\cite{chiang2011exploiting,leskovec2010signed}, sign prediction \cite{hu2019sparse}, community detection\cite{hu2019sparse} and visualization\cite{kunegis2010spectral}. The network embedding usually finds a mapping function that converts each node to a low-dimensional latent representation in a network. However, most of the works of existing graph embedding algorithms have been used for unsigned homogeneous and heterogeneous graphs. Homogeneous graphs \cite{grover2016node2vec,salehi2017properties,schlotterer2019investigating,tang2015line} are composed of similar nodes and edges whereas heterogeneous graphs known as heterogeneous information network (HIN) \cite{sun2013mining,dong2017metapath2vec,fu2017hin2vec,tang2015pte} are capable of composing different entities (i.e. nodes and relations), have become ubiquitous in real-world scenario like social network, recommendations system.

The first paper \cite{yuan2017sne} gave attention to sign network embedding and built a log-bilinear model named SNE which used node representations of nodes and edge sign information along a given path and an objective function for optimization during the node embedding learning process. As social theories are different in signed graphs than in unsigned graphs, CSNE \cite{mara2020csne} introduced a probabilistic model that  integrates structural information and signs information separately. Not only linear modeling functions have been used for network embedding but also deep learning-based techniques also have been used for network embedding. SiNE \cite{wang2017signed} optimizes an objective function mainly using the social theories. SiNE builds a  deep learning framework by learning node representation. SGCN, SiGAT, SNEA \cite{derr2018signed,huang2019signed,li2020learning} utilizes balance theory \cite{cartwright1956structural} and aggregate negative link information from graph data and introduce an optimization function to learn node representation. Beyond link prediction, a sign of a link has also been predicted. DNS-SBP \cite{shen2018deep} applied a deep network model that employs a semi-supervised auto-encoder focusing on negative links preserving structural balance theory in real-world signed networks. StEM\cite{rahaman2018method} used decision boundary for node representation to separate friends nodes to foes nodes and shows higher quality representation over distance-based ranking approach. Computation cost is also an important concern in graph embedding, focusing on training time and optimization techniques, SIDE \cite{kim2018side} used a general network embedding method to model a network that consists of multiple steps of relationships to represent sign and direction. 

All of the above-mentioned works focused on homogeneous signed networks which only consist of one type of node and the nodes are specialized to apply social balanced theory and ignore link and node heterogeneity. These models cannot be directly applied to heterogeneous signed graphs. However, in real-world scenarios, signed graphs are not always homogeneous. Recently, researchers put their attention to signed heterogeneous graph cases where there can be multiple node types e.g. in social network analysis nodes can be user nodes and topics. SHINE \cite{wang2018shine} model uses three autoencoders to extract short and dense embedding from three different types of networks aggregate these embeddings into final heterogeneous embedding and leverage such information for sign link prediction. SiHet \cite{rizi2020signed} proposed some improvements over SHINE investigating how to join encode sentiment and social networks into low-dimensional representation preserving sign and network structure while retaining a linear time complexity. ComPath \cite{dhelim2020compath} introduced a link prediction algorithm to predict unknown links in heterogeneous sign networks. A similarity check proximity function  measures the similarity between users and an adaptive clique relaxation technique detects communities with common interests. 

In comparison with signed networks, most of the research surveys have focused on unsigned network analysis which includes both homogeneous\cite{cai2018comprehensive,goyal2018graph} and heterogeneous graph\cite{wang2020survey} for various tasks including link prediction, node classification. In this article, we present a comprehensive review of current graph embedding methods that have been proposed for signed graphs in real-world applications.

We will thoroughly survey the existing works on signed graph embedding methods and applications. We will also avail the dataset that has been used for most of the research work and available open-source code. In short, (1) we will explore the recent techniques of sign graph embedding covering both homogeneous and heterogeneous graphs analyzing their pros and cons (2) We will discuss the existing methods that have been applied in real-world applications (3) we will summarize available benchmark datasets and open-source code for future research scope in this field  (4) we will explore scholar data co-author network and analyze the  features of the network (5) we will discuss additional challenges and new research directions based on the existing work.
There has not been much survey work done on signed network embedding. \cite{tang2016survey} provided a brief survey of basic concepts, and unique principles of signed network embedding, and classified mining tasks of social media analysis into node-oriented, link-oriented, and application-oriented tasks. We make our contribution to this work is summarized below:

\begin{itemize}
\item We first discuss theories of  sign graph embedding  and then provide a comprehensive survey of existing homogeneous and heterogeneous graph embedding methods. We will categorize the learning process based on the information that has been used to address particular challenges for signed graphs.

\item We summarize benchmark datasets and open-source code with a given description to facilitate future research work in this field.
\item We explore additional challenges and the future directions of signed graph embedding.

\end{itemize}

\begin{figure*}
\centering
\begin{minipage}[b]{.33\textwidth}
\includegraphics[width = 5cm,height=2cm]{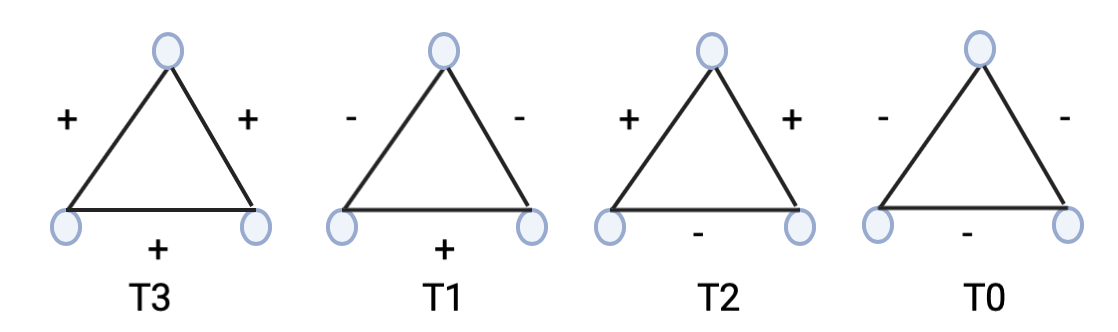}
\caption{\raggedright According to \cite{leskovec2010signed}, triads with odd positive
edges are balanced (T3, T1), and triads with even positive edges (T2, T0) are unbalanced.}\label{label-a}

\end{minipage}\qquad
\begin{minipage}[b]{.4\textwidth}
\includegraphics[width = 9cm,height=2.5cm]{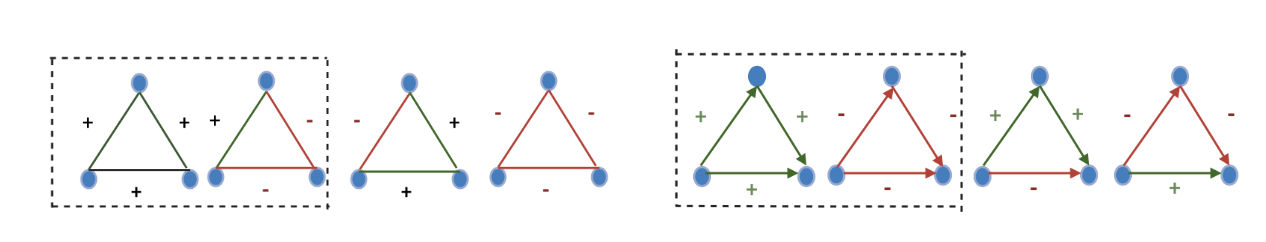}
\centering{(a) Balance Theory (b) Status Theory}
\caption{\raggedright Sociological theories on directed signed triads. In both (a) and (b), the first two triads are balanced and the last two triads are unbalanced.}\label{label-b}
\end{minipage}

\end{figure*}
\section{\textbf{\LARGE Preliminary}}
In this section, we briefly describe the basic concepts, principles, and properties of signed networks which are related but distinct from unsigned networks. Then, we elaborate on the challenges brought by signed networks compared with unsigned networks. 

\subsection{\textbf{Signed Networks}}

In signed network, considering a graph  $G^{+-} = (G^{+},G^{-})$. Signed undirected networks are represented by G = (V,E) where V = $\{v_{1},v_{2},v_{3},...,v_{n}\}$ denotes vertices and = $E^+ ,E^-$, we denote $E^+$ as positive links and $E^-$ as negative links. Here $E \subseteq \binom{V}{2}$.
Each node denotes an entity and each edge denotes a relation between two entities. Edges can be described as positive and negative relations. $E^{+-}  \rightarrow (+1, -1)$ is used to map edges to their respective signs. To present both positive and negative links into one adjacency matrix $A \in R^{NxN}$, where $A_{ij} = 1, A_{ij} = -1 and A_{ij} = 0$ denote positive, negative and missing links from $v_i$ to $v_j$, respectively. $A^{+-}$ = $A^+ + A^-$, here the positive  and  negative adjacency matrices are considered separately to represent the signed network.     

\subsection{\textbf{Theories of Signed Networks}}
To analyze a large-scale dataset from social phenomena, the theories of signed networks have been applied to social psychology. It helps to reason out the impact of different kinds of relationships across the applications. We analyze signed networks using two different theories: structural balance theory and status theory and a comparison of these two theories.

\subsubsection{\textbf{(i) Structural Balance Theory}}

According to Cartwright and Harary \cite{cartwright1956structural} structural balance in graph theoric language considers possible which ways in triangles on four individuals can be signed. In \cite{leskovec2010signed}, the theory is described with four triangle models $T_0 ,T_1,T_2,T_3$. Here, triangle models are defined as  $T_3$: three mutual friends, $T_1$: Two friends with a common enemy, $T_2$: Two enemies with a common friend, or $T_0$: Three mutual enemies. Here, balanced triangle $T_3$ is based on the principle that "a friend of my friend is my friend" and  $T_1$ triangle with two negatives and one positive edge can be described in three ways: "a friend of my enemy is my enemy", "an enemy of my enemy is my friend", and "an enemy of my friend is my enemy".

\subsubsection{\textbf{(ii) Status Theory}}

Balance theory signifies more about likes and dislikes, however, another meaning can be interpreted from these triangles apart from likes/dislikes. The status theory uses the positive and negative labels as an interpretation of status within each other. For example, a positive link from A to B means "A likes B" and can also be interpreted as "B has a higher status than A". Similarly, a negative link from A to B can be interpreted as "B has a lower status than A". 

Based on the status theory, here we consider a positive directed link indicates that the source of the link views the target as having higher status, and a negative directed link indicates that the source is viewed as having lower status than the target.   

\subsubsection{\textbf{(iii) Comparison of two theories}}

To understand the difference between these two theories, we look into a situation where a positive link is from user A to user B and a positive link is from user B to user C. According to balance theory, we expect that a link from C to A should be a positive link based on the principle "a friend of my friend is my friend". However, according to status theory, A links to B positively means "A regards B as having higher status" and B links to C positively means "B regards C as having higher status". According to status theory a link from C to A inclined to link negative to A as "C regards A as having lower status "

\begin{figure}[tbh]
\centering
\includegraphics[width=6cm]{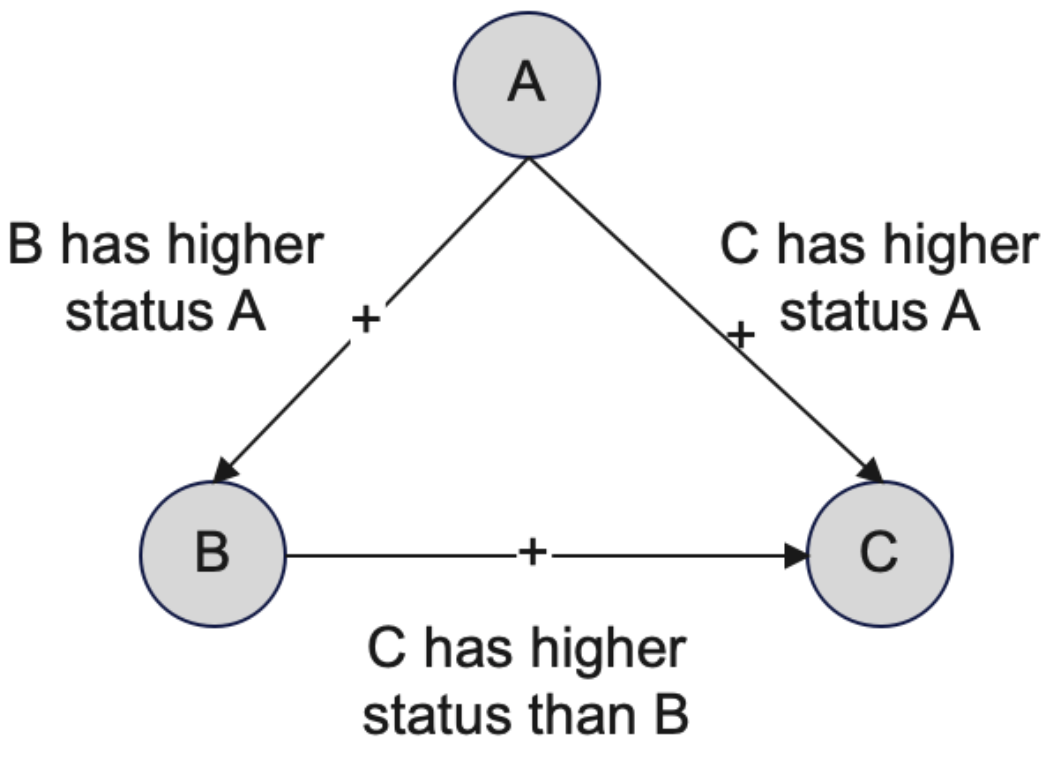}
\centering {\caption{\raggedright Illustration of Status Theory}}
\end{figure}

The prediction of balance theory for certain types of triads is represented differently than status theory. 

\begin{figure*}
\begin{minipage}[b]{.3\textwidth}
\includegraphics[width = 5cm,height=7cm]{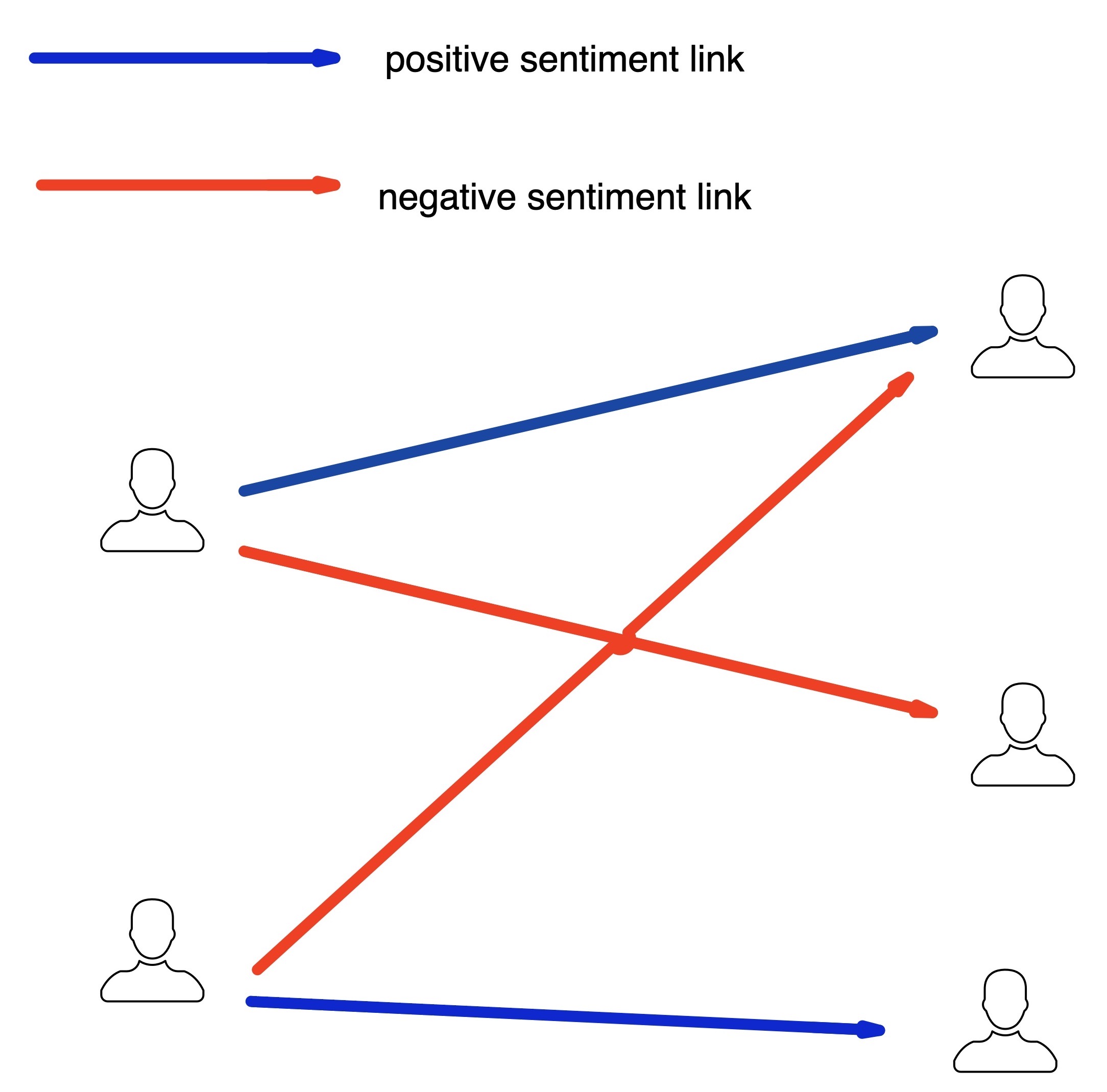}
\caption{\footnotesize Sentiment network}\label{label-a}
\end{minipage}\qquad
\begin{minipage}[b]{.3\textwidth}
\includegraphics[width = 4cm,height=6cm]{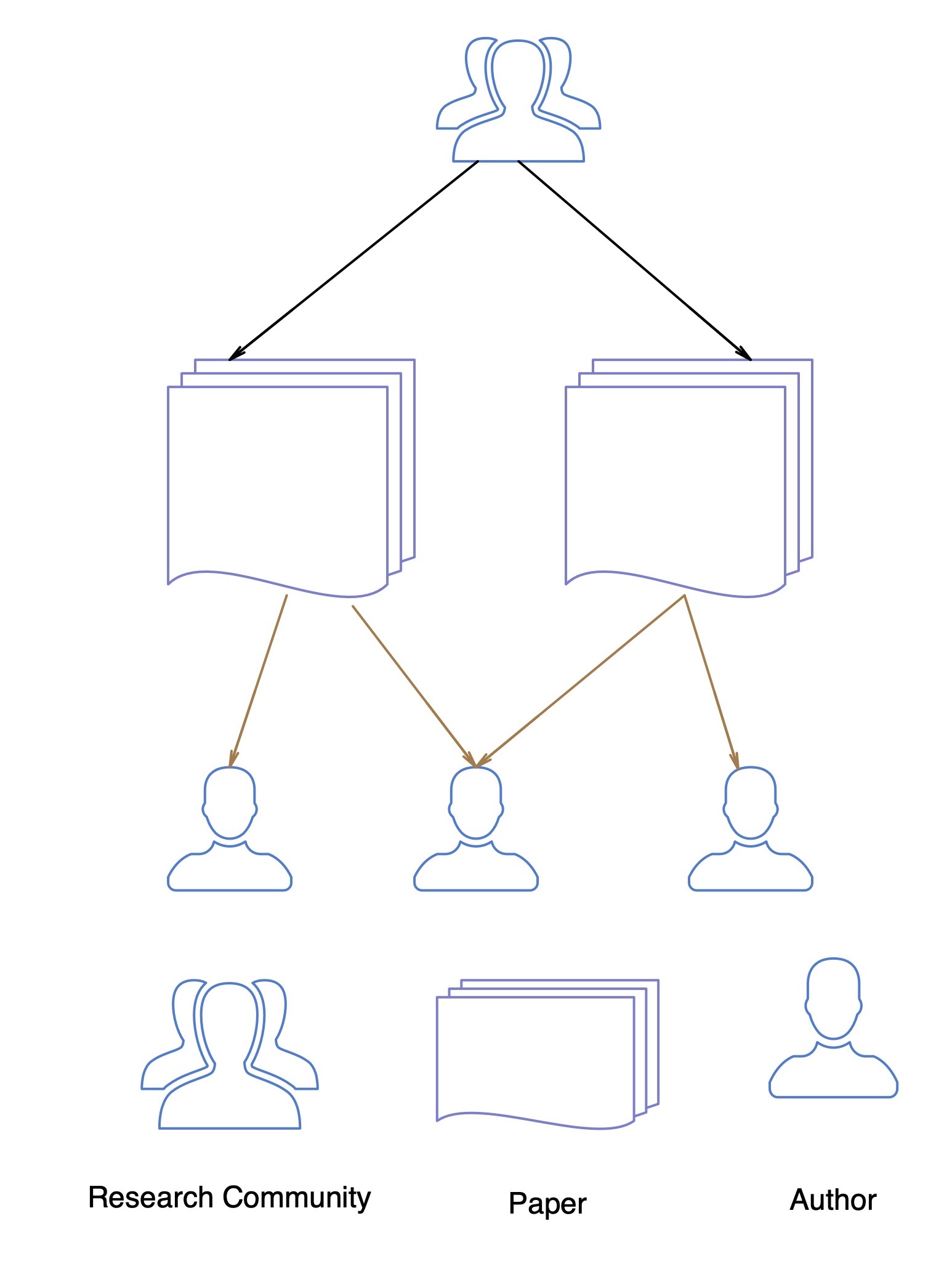}
\caption{\footnotesize {Example of Heterogeneous network}}\label{label-b}

\end{minipage}
\begin{minipage}[b]{.3\textwidth}
\includegraphics[width = 4cm,height=7cm]{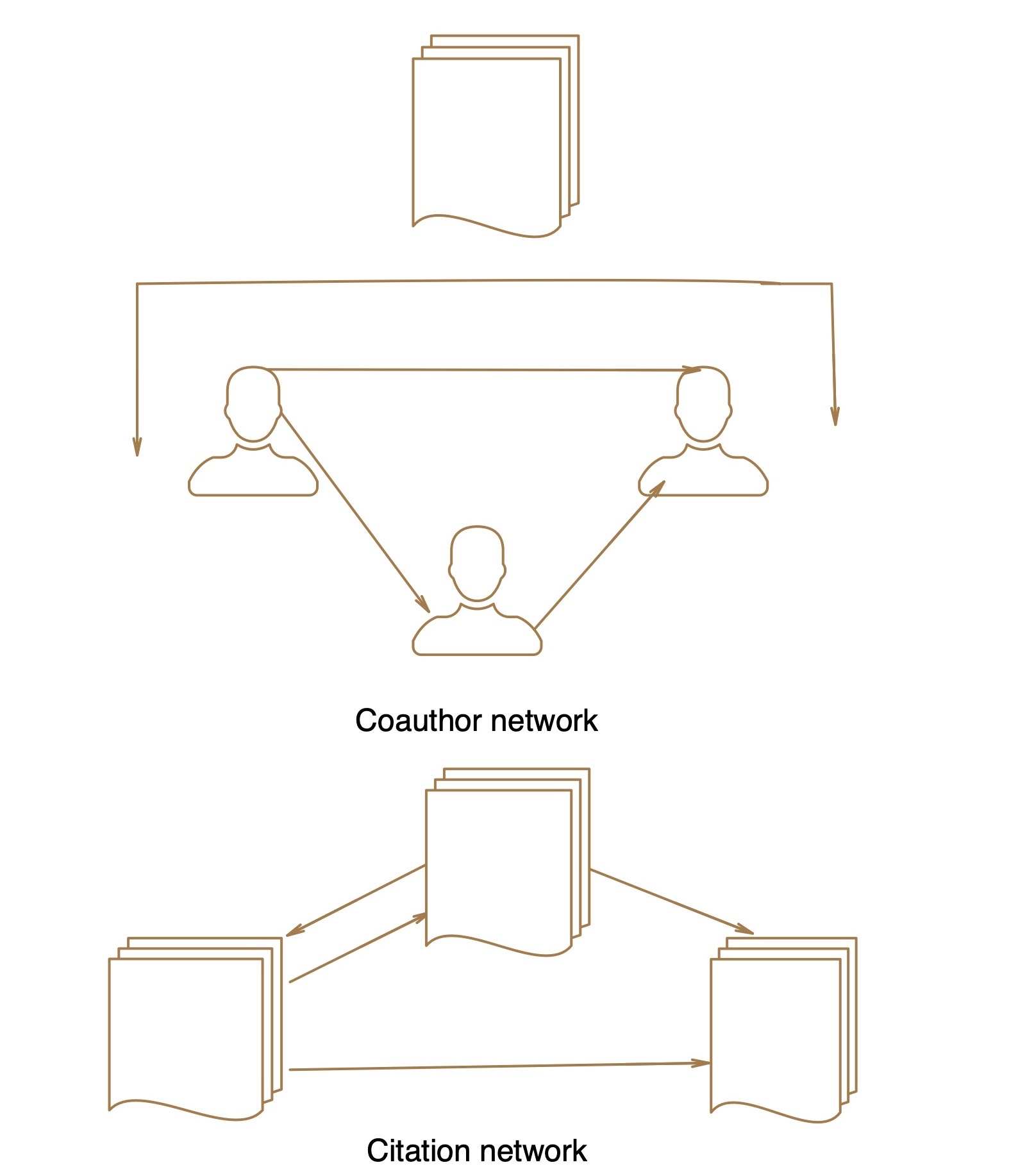}
\caption{\footnotesize Citation network}\label{label-b}

\end{minipage}
\end{figure*}

\subsection{\textbf{Network Analysis}}
In graph theory, network analysis is  an application that is concerned with analyzing relational data. The network analysis addresses the importance of actors in a given network and the concentration of the network. It also helps to understand the properties of the network such as size, density, degree, and diameter. One of the most common interests of structural analysts is in the substructures that exist in the network. The closed community of actors or "sub-groups" is an important aspect of social structure. It is essential to see if there are particular actors that appear to play network roles. 

\subsection{\textbf{Signed Network Embedding}}
A signed network is defined as $G = (V, E^+, E^-$), where $V$  is the set of vertices and $E^+$,$E^-$ are the sets of positive and negative edges, respectively. Signed graph embedding aims to learn a function $ g: V \rightarrow \mathbb{R}^d$ that embeds nodes to $d$ -dimensional real-valued vectors. The representations are denoted as $X = \{x_1, x_2, x_3, \ldots, x_n\} \in \mathbb{R}^{n \times d} $, where $x_i$ is the embedding corresponding to node $i$. 

Unlike unsigned graphs, in signed network embedding, we additionally require function $g: E \rightarrow \mathbb{R}^d$  to keep negative edges at a greater distance in the embedding space compared to positive edges. Here, $E$ represents the links to map pairs $\{i, j\} \in E^-$ to more distant representations $x_i$ and $x_j$ than pairs $\{i, j\} \in E^+$ . Additionally, unlinked nodes are not used for embedding learning in a signed network because their distance is unimportant.

\section{\textbf{\LARGE Aspects of Signed Network}}
Different from unsigned graph embedding \cite{goyal2018graph}, signed graph embedding methods have been designed for networks with different edge types and the edges are mostly means to relationships. Different types of e-commerce  and social media sites such as Amazon, Facebook, Twitter, and Epinions can be represented as a signed network where positive represents trust, and agreement while negative ones may show distrust and disagreement. Different edge types in the signed graph impose more challenges in  the sections illustrated below.

\subsubsection{\textbf{Structure}}

In an unsigned network, the links are considered as positive. The algorithms developed for unsigned networks such as PageRank \cite{chung2014brief} require all links to be positive. Spectral clustering \cite{von2007tutorial} algorithms for unsigned networks are not applicable to signed networks. Some social theories for unsigned networks such as homophily may not be applicable to signed networks while status theory and balance theory are only applicable to signed networks. Previous survey works focused on heterogeneous network embedding discarding the sign information in the network \cite{wang2020survey,yang2020heterogeneous,shi2016survey}. To address the sign information, current research has been putting efforts towards structural theory and other novel ideas to capture the sign link information.

\subsubsection{\textbf{Attributes}}
Most of the signed networks contain similar types of nodes for example users with different relationships as positive and negative. However, in social media, signed networks may not be confined to only positive and negative links along with the same types of nodes as users. For example, in social media, there can be sentiment relationship (positive, negative), social relation (follower, followee), (interest, disinterest)\cite{dhelim2020compath,wang2018shine}. In heterogeneous graph, the attributes of nodes may contain important information, for example, gender, and location. There can exist different types of nodes as well e.g. users, topics nodes, topics community \cite{wang2018shine}. The heterogeneity of nodes and edges brings another dimension to the signed network and effectively fusing the information of attributes poses another challenge.  
 
\subsubsection{\textbf{Applications}}

Signed graphs are closely related to real-world applications. Different methods have been explored to capture structures of networks for different applications. Homogeneous signed networks are comparatively simpler than heterogeneous signed networks. In that case, for better prediction, node, and link  information need to be carefully encoded \cite{dhelim2020compath,wang2018shine,rizi2020signed}.  


 \section{\textbf{ \LARGE Method Taxanomy}}

In the signed graphs, links carry relationships between nodes. There are different types of approaches that have been applied to embed signed graphs. These approaches include similarity-based, structure-based, and attribute-assisted methods. Based on the methods that have been proposed so far in this field, we have categorized the approaches that have been used for signed graph embedding: 
(1) Log-bilinear model-based SG embedding: The method captures both the node's path and sign information. 
(2) Proximity measure-based SG embedding: The methods capture node's path and edge information to predict target node based on its path and measure similarity between representations. 
(3) Path-based SG embedding: The methods use random walk to capture node's information along a path. 
(4) Neural network-based SG embedding: Neural network-based methods use random walk, proximity measure, and structure preserved techniques for signed graph embedding. 


\begin{table}[h]
\caption{Table of symbols} 
\centering
\begin{tabular}{c| r}
\hline\hline
Symbol&\multicolumn{1}{c}{Meaning} \\ [0.5ex]
\hline 
V & Set of nodes \\
u,v & nodes\\
q & type of edge \\[1ex] 
$w^{u}$ & low dimensional vector representation of $u$ \\
$w^{v}$ & low dimensional vector representation of $v$ \\
$w^{k}$ & low dimensional vector representation of $k$ \\
$w^{n}$ & weights of the $n$-th layer \\
$b^{n}$ & bias of the $n$-th layer \\
$W^{out}$ & out embedding matrix \\
$W^{in}$ & in embedding matrix \\
$b^{in,+}$, $b^{in,-}$ & positive/negative in-bias vectors \\
$b^{out,+}$, $b^{out,-}$ & positive/negative out-bias vectors \\
$\sigma$ & sigmoid function \\
$W^{B}$ & Balanced node sets \\
$W^{U}$ & Unbalanced node sets \\
$h^{B}$ & Hidden representation of balanced node sets \\
$h^{U}$ & Hidden representation of unbalanced node sets \\
\hline
\end{tabular}
\label{tab:hresult}
\end{table}

\subsection{\textbf{\large Log-bilinear model based SG embedding}} 
Unlike unsigned networks, signed network embedding needs to capture both node and sign information. For example, unsigned networks based on the skip-gram model only capture node information and cannot capture the sign information. SNE~\cite{yuan2017sne} first proposed a graph representation method for signed networks, which uses node representations of all nodes along a given path and captures sign information along the path using a log-bilinear model. The log-bilinear model predicts the target node given its predecessors along a path. 

To embed a signed network, a signed type vector ($c_i$) ($c_+ \in \mathbb{R}^d$ and $c_- \in \mathbb{R}^d$) is defined to represent signed type (positive, negative) edges. To compute the target node embedding, SNE combines source embeddings ($S_{u_i}$) of all nodes along a given path $h = [u_1, u_2, u_3, \ldots, u_l, v]$ with corresponding signed type vectors ($c_i$) by performing an element-wise multiplication. A score function calculates the similarity between the actual ($\hat{v}_h$) representation and predicted ($v'_v$) representation of the target node $v$ (eq (1)). It then computes the conditional likelihood of the target node given the path of nodes $h$ and their edge types $q$ based on the softmax function (eq (2)).


\begin{equation} \label{eu_eqn}
Score(v,h) = {\hat{v}^T_{h}v^{'_v} + b_v}
\end{equation}

\begin{equation} \label{eu_eqn}
p(v|h,q;\theta) =  \frac{exp(Score(v,h))}{\sum_{v^{'}\in V}^{} exp(Score(v^{'},h))}
\end{equation}

\subsection{\textbf{\large Proximity measure based SG embedding}} 
SiHet \cite{rizi2020signed} use joint probability between vertex u and v integrating with sentiment edge. Join probability is a generic way to model proximity between vertex u and v in the network is described in LINE \cite{tang2015line} as follows: 

\begin{equation} \label{eqno} \begin{aligned}
p_r(u,v) =  \sigma(w_u . w_v)
\end{aligned}
\end{equation}
The joint probability value increases to 1 for positive sentiment and get smaller approaching to 0 for negative sentiment because of the inner product term, $w_u.w_v$

\begin{equation} \label{eqno} \begin{aligned}
p_s(u,v) =  \sigma(s_{uv * (w_u . w_v)})
\end{aligned}
\end{equation}
\begin{equation} \label{eqno} \begin{aligned}
O_s = - \sum_{ (u,v) \in S^+}^{} [\log p_s(u,v) + \sum_{j=1}^{}\log p_s(u,v_j)]
\end{aligned}
\end{equation}

SiHet designed an objective function (eq(5)) that randomly chooses k negative signs for every positive edge(u,v). The minimization of the objective function embedding vectors gives high similarity values to the nodes that are positively connected and low similarity values to the negatively connected nodes. SiHet does the heterogeneous network embedding by combining a social network embedding along with sentiment network embedding for link prediction and node recommendation.
The social network embedding preserves the local neighborhood structure. For a social edge (u,z), first-order proximity is preserved by maximizing joint probability between two vertices. 

\begin{equation} \label{eqno} \begin{aligned}
O_r = - \sum_{ (u,z) \in R} \log p_r(u,z)
\end{aligned}
\end{equation}
\begin{equation} \label{eqno} \begin{aligned}
O_{SiHet} = - O_s + O_r
\end{aligned}
\end{equation}
 The collectively combined objective function of two networks is achieved by minimizing objective function (eq 7)

The proximity function is also used in ComPath \cite{dhelim2020compath} to measure similarity among users and topics to construct heterogeneous graphs. Based on that, the author proposed to detect communities of similar interest and cluster semantically similar topics. The similarity measurement has two steps: User Similarity and Topics Similarity. To compute user similarity, ComPath proposed a similarity function relative proximity function RPF which computes (i) SimP: similarity of user interest (ii) SimN: similarity of user disinterest (iii) intersection between interest and disinterest. The similarity function measures similarity between intersection of user interest and intersection of disinterest. 
\begin{equation}  \label{eqno}
\begin{aligned}
RPF(u_x, u_y) &= \text{SimP}(U_x, U_y) + \text{SimN}(U_x, U_y) \\
&\quad - \left(\frac{|P_x \cup N_y|}{|P_x \cup N_y|} + \frac{|N_x \cup P_y|}{|N_x \cup P_y|}\right)
\end{aligned}
\end{equation}


To compute the user interest similarity, eqn(8) computes the RPF by summing (SimP+SimN) along with Jaccard similarity index of the mutual interest/disinterest intersection. 
To compute topic similarity, ComPath uses a topic similarity objective function that measures the similarity between two topics: (i) SimTP: the similarity of two topics common interested users (i) SimTN: the similarity of two topics common disinterested users. 

\begin{equation} \label{eqno} \begin{aligned}
SimT(t_x,t_y) = (SimTP(t_x,t_y)) + (SimTN(t_x,t_y))
\end{aligned}
\end{equation}

The overall topic similarity eq(9) is the sum of the similarity of topics commonly interested users and the sum of topics disinterested users.
In summary, heterogeneous signed graph embedding methods can be divided into two categories: (1) Explicitly preserve the proximity of links [52], [53]; (2) Preserve the proximity of nodes, which utilizes the information of links implicitly [17], [50], [18]. Both of the methods use the first-order information of heterogeneous signed graphs.

\subsection{\textbf{\large{Path-based SG embedding}}}
Other methods have also been proposed to embed signed directed networks based on random walks. SIDE \cite{kim2018side} is a general network embedding model that represents both the sign and direction of edges in the embedding space. SIDE aggregates sign and path information according to balance theory and direction following topological order.  The method is based on truncated random walks and computes a likelihood formulation for signed directed connection. \\ 
Random walk-based methods usually create multiple truncated node sequences which reveal proximity among nodes. Two nodes are closer to each other in random walk sequences when the links among them have high weight values and shorter paths. 

\begin{figure}[tbh]
\includegraphics[width=6cm]{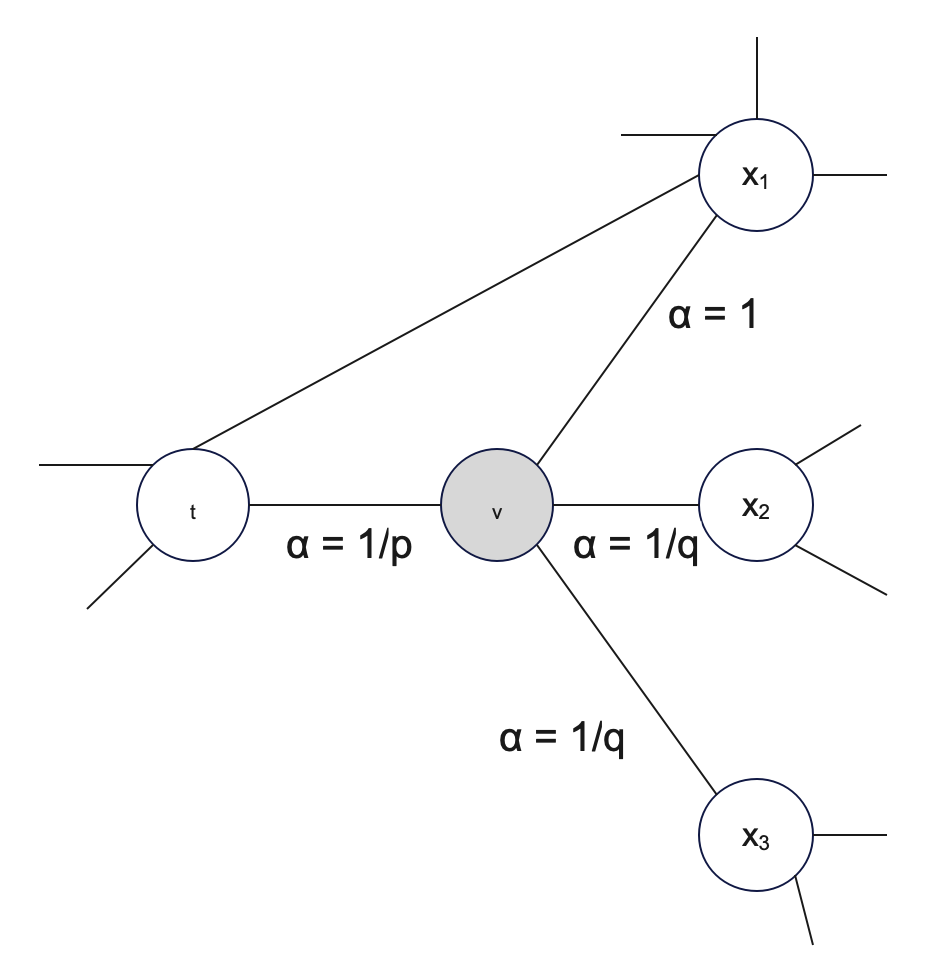}
\caption{\raggedright The random walk procedure in node2vec \cite{grover2016node2vec}. The walk starts from node t to v and on the next step transit to the next nodes from node v. The transition depends on the search biases labeled indicated over edge.}
\end{figure}

A random walk is simulated with a fixed length of L from a given source node (Fig 7). Let $x_i$ denote the $i_{th}$ node in the walk, starting with $x_0$ = u. The random walks would be to sample the next node based on bias term $\alpha$ on the static edge weights. Nodes $c_i$ are generated by the following distribution:
 
\begin{equation} \label{eqno} \begin{aligned}
P(x_i = u | x_{i-1} = v) = 
    \begin{cases}
        \frac{\pi_{vu}}{Z}  \hspace{0.5cm}  \text{if (u,v)} \in  0   \\
        \text{0  \hspace{0.5cm} otherwise} 
    \end{cases} 
\end{aligned}
\end{equation}
Here $\pi_{vx}$ is the unnormalized transition probability between nodes v and x, and Z is the normalizing constant. \\
In SIDE, the first step generates multiple truncated random walks for directed edges until the required length is fulfilled. In the directed network, random walk follows the topological order. 
In the second step, SIDE learns vector representation using Skip-gram with negative sampling (SGNS)\cite{mikolov2013distributed,mikolov2013efficient}. SGNS model predicts context words for a given central word. It formulates maximum likelihood of co-occurrence of node pairs.  The likelihood predicts whether a pair of nodes co-occur or not. The likelihood of u linking to v is formulated as follows:

\begin{equation} \label{eqno} \begin{aligned}
P(u,v) = \sigma(W_u,W_v') = \frac{1}{1+exp(-W_u.W_v')}
\end{aligned}
\end{equation}
Here $\sigma$ is a sigmoid function and $W_u$ and $W_v'$ are the "target" and "neighborhood" embedding vectors, respectively. The larger product value $W_u.W_v'$ has a higher likelihood of co-occurrence. The negative maximum log-likelihood formulation links embedding vectors to likelihood of sampled pairs. The objective function of SIDE is defined below: 

\begin{equation} \label{eqno} \begin{aligned}
J = \sum_{(u,v) \in \textit{D}} ([-log P(u,v)) + \sum_{n}^{j = 1} - logP(u,v_j')] 
\end{aligned}
\end{equation}

For each pair from set \textit{D} with aggregated sign and direction, noise samples $v_j'$ were randomly selected to form noise pairs. 


The likelihood function is defined as below:

\begin{equation} \label{eqno} \begin{aligned}
  P(u,v) =
    \begin{cases}
        \sigma(W_u^{out}.W_v^{in}+b_u^{out,^+}+b_v^{in,+}) \hspace{0.1cm} \text{ if sign(u,v) $>$ 0}   \\
        \sigma(-W_u^{out}.W_v^{in}+b_{u}^{out,^-}+b_{v}^{in,-}) \hspace{0.1cm} \text{if sign(u,v) $<$ 0}   \\
        \sigma(-W_u^{out}.W_v^{in})  \hspace{0.5cm} \text{if v is a noise}
    \end{cases} 
\end{aligned}
\end{equation}

To estimate the sign of a link, the first component of the likelihood function increases as the inner product increases for the positively connected nodes. On the other hand, for negative and noise pairs, the likelihood value increases when inner product similarity is lower. The maximum likelihood keeps the balance of positive push and negative pull which actually keeps nodes with similar signed neighborhoods close and opposite signed neighborhood structures far apart in embedding vector space. \\
In equation (13), the positive, negative, and noise pairs likelihood function is calculated in first, second, and third equations respectively.
Sparse-node2vec \cite{hu2019sparse} model produces node sequences using biased random walk based on structural information. The model represents community structure and neighbor relationships. The two kinds of network structure information described in the paper are: neighbor relationships which are based on structural balance theory and common neighbor relationships is same as community structure which involves a friend set and an enemy set. The same community consists of more positive links and more negative links between different communities. 
SNE \cite{yuan2017sne} algorithm also adopted a log-bilinear model to predict target node based on path generated by a truncated random walk. It first randomly initializes node embeddings and then uses uniform random walk to generate corpus and repeat the process t times.

\subsection{\textbf{\large{Neural network based SG embedding}}}
\subsubsection{\textbf{\large{Proximity measure}}}

On the previous section, we have seen that the representation learning involved factorizing a matrix encoding of the graph. Some of the models applied a neural embedding model to solve an auxiliary task related to the observed topology of the graph. Though neural net has been used for unsigned graphs, the same model will not apply to signed graphs. In representation learning a node in vector space resides close to the representation of its neighbors but in the signed graph nodes placing close proximity with foes is undesirable. StEM \cite{rahaman2018method} first introduced neural embedding that learns the representation of signed networks.

StEM learns a decision boundary to separate each node's friends from its foes. A decision boundary separates u's friends from u's foes. The relative similarity of two friend nodes will be positive otherwise negative. The similarity measure function is defined as: 

\begin{equation} \label{eqno} \begin{aligned}
\phi(w^u; \beta) M_2 \phi(w^v; \beta_v)^T + b_2
\end{aligned}
\end{equation}

Here, $\phi(w^u;\beta)M_2$ interpreted as the decision boundary, 
$M_2$ the learnable matrix of weights and  $\phi:\mathbb{R}^d \longrightarrow \mathbb{R}^d$ is a parameterizable activation function and $\beta$ is learnable parameter of $\phi$. Here the $\phi$ is defined as single channeled parameterized leaky rectified linear unit.

The representation function f is defined as \begin{equation} f: V \longrightarrow \mathbb{R}^d. \end{equation}

\begin{equation} \label{eqno} \begin{aligned}
f(u) = (M_1g(u))^T + b_1)^T
\end{aligned}
\end{equation}

\begin{equation} \label{eqno} \begin{aligned}
g(u) = W_u
\end{aligned}
\end{equation}

Here, M1 $\in \mathbb{R}^{d \times 2d}$, W $\in \mathbb{R}^{N \times 2d}$ and b1 $\in \mathbb{R}$ are all learnable parameters. A feedforward network accepts indices of two nodes u and v, and outputs the probability of a positive edge existing from u to v.
$W_u$ is the $u^th$ row of matrix. The forward pass is described below:

\begin{equation} \label{eqno} \begin{aligned}
\begin{split}
r_u = W_u\\
x_u = (M_1r_u^T + b_1)^T\\
q_u = PLeakyRELU(x_u,\beta)\\
z = q_uM_2q_u^T + b_2
\end{split}
\end{aligned}
\end{equation}

To calculate the loss, the loss function is described as 

\begin{equation} \label{eqno} \begin{aligned}
\begin{split}
z = q_uM_2q_u^T + b_2
\end{split}
\end{aligned}
\end{equation}

This paper \cite{song2018learning} also has used neural networks for signed graph embedding. The model keeps both bode and edge patterns. It uses second-order node proximity for signed networks. An objective function preserves the node and edge information. Node proximity is satisfied when two nodes are similar in sign network and have similar sign context.

\begin{figure}[tbh!]
\includegraphics[width=8cm]{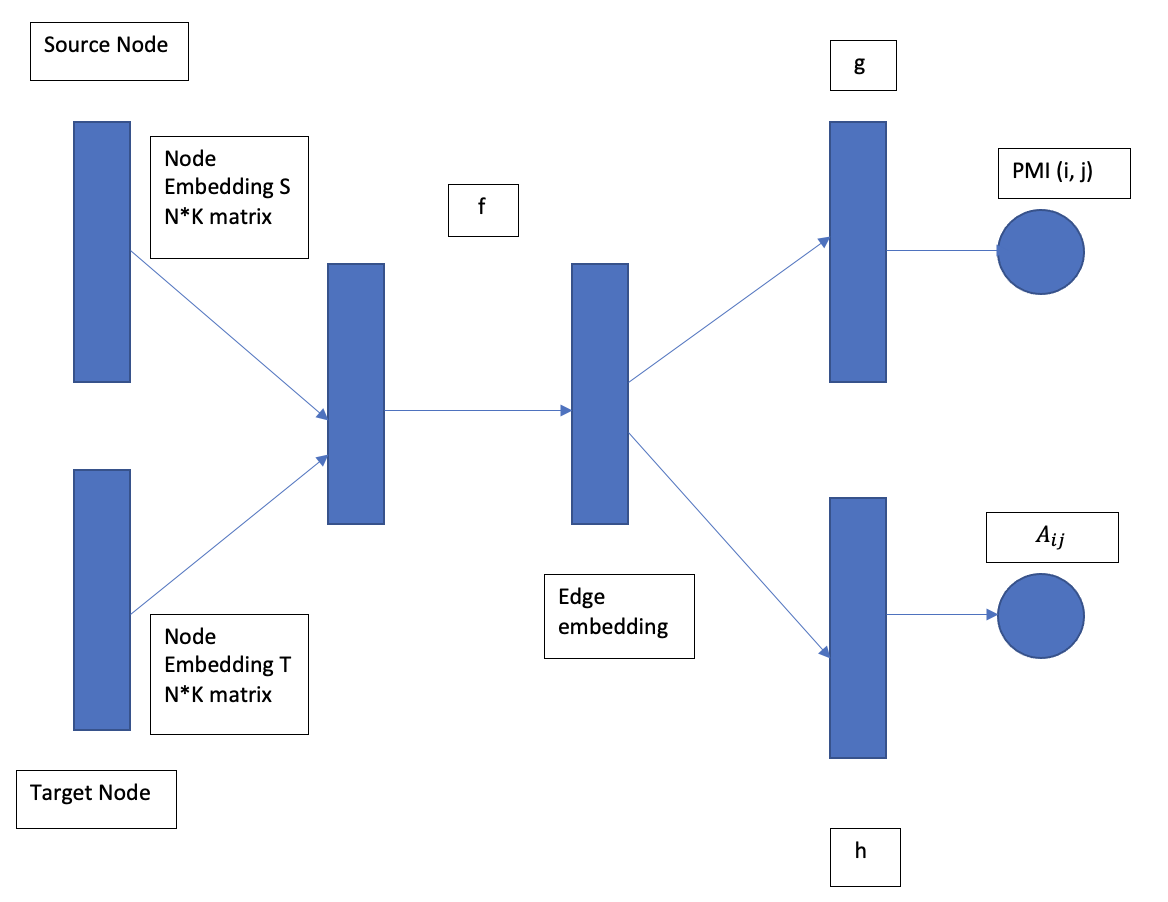}

\caption{\raggedright The framework nSNE \cite{song2018learning}   has 4 hidden layers. The network contains the embeddings of source nodes target nodes and observed edges. The mapping functions in nSNE, i.e., f, g, and h generate output of this neural network. The embeddings of nodes and mappings are learned by the back-propagation algorithm.}
\end{figure}

Second-order proximity considers mutual dependence of two nodes. The neural network signed network (nSNE) embedding learns node vectors and mapping function via neural network approach, The model used a Multi-Layer Perceptron to model the mapping function. The embedding of nodes and edges is integrated together to learn embedding and mapping functions simultaneously via backpropagation. The network has four types of layers: 
(i) input component: (ii) node-to-edge mapping component (iii) edge-to-weight task component (iv) edge-to-sign task component. 

The input components consist of two N-dimensional one-hot vectors which represent source and target node of observed edge. Two fully connected layers create function for node-to-edge mapping. The first layer of this component inputs the element-wise product of two vectors. It contains K neurons, and the weight matrix of this layer i.e., $S^{N \times K}$ and $T^{N \times K}$ represents source and target vectors of all the nodes.  The edge-to-weight task component models the relationship between edge vector and its weight or PMI (Pointwise Mutual Information). Mean Squared Error (MSE) as loss function calculates the difference between predicted and real PMI. The edge-to-sign task components learn the relationship between edge vector and the sign vector. The last layer  uses single neutron with sigmoid function and Binary Cross Entropy (BCE) to measure distance between real and predicted PMI.

\subsubsection{\textbf{\large Structure preserved SG Embedding}} 

Signed graph embedding methods capture properties of signed graphs. One of the important social properties of signed graph is structural balance theory. According to balance theory\cite{harary1953notion}, a signed network is balanced if and only if within-cluster all edges are positive and between clusters all are negative. An extended structural balanced theory proposed in \cite{cygan2012sitting} states that nodes with positive links should sit closer than those connected with negative links. In this section, we review the typical signed graph embedding methods based on the graph structures. 
These methodical structures are the fundamental techniques of signed graph embedding, which are capable of capturing structural information of signed networks. In the following, we will discuss elaborately the typical structure-preserved signed graph embedding methods based on these two types of structures and then discuss their improvements of one over another. 
\\
SiNE\cite{wang2017signed} model built based on the objective function for signed network embedding by extended structural balance theory to learn a good representation of signed network. The objective function is defined below:

\begin{equation} \label{eqno} 
\begin{aligned}
\min_{(X,x_0,\theta)} \frac{1}{C} \left[\sum_{(w^u,w^v,w^k) \in P} \max\left(0, f(w^u,w^k) + \delta - f(w^u,w^v)\right)\right] + \\ \text{regularization term}
\end{aligned}
\end{equation}

Here $\theta$ is the similarity function, training data is P, and the regularization term is used to avoid overfitting.

The aim of the object function is to learn node representation and measure node similarity in a sign network. SiNE designed a deep learning non-linear function that consists of 2 hidden layers and is generalized into N layers. The input of the framework is the set of triplets extracted from the signed network as ($v_i,v_j,v_k$) with $e_ij$ = 1 and $e_ik$ = -1. For N deep networks, the input of the n-th layer is the output of (n-1)th layer. The parameters of the deep net are initialized by a uniform sampling. Here, weights of the hidden layers are initialized using uniform sampling and embedding vectors initialized as zero matrices.

The output of the first layer $z^11$ and $z^12$ are inputs of the second hidden layer. 
\begin{equation} \label{eqno} \begin{aligned}
\begin{split}
z^{11} = tanh(W^{11}w^u + W^{12}w^v + b^1) \\
z^{12} = tanh(W^{11}w^u + W^{12}w^k + b^1)
\end{split}
\end{aligned}
\end{equation}

The output of the n-th layer:
\begin{equation} \label{eqno} \begin{aligned}
\begin{split}
z^{n1} = tanh(W^nz(n-1)1 + b^n) \\
z^{n1} = tanh(W^nz(n-1)2 + b^n) \\
\end{split}
\end{aligned}
\end{equation}

The output of the N-th layer:
\begin{equation} \label{eqno} \begin{aligned}
\begin{split}
f(x_i,x_j) = tanh(w^Tz^N1 + b) \\
f(x_i,x_k) = tanh(w^Tz^N2 + b)
\end{split}
\end{aligned}
\end{equation}

\begin{figure}[tbh]
\includegraphics[width=9cm]{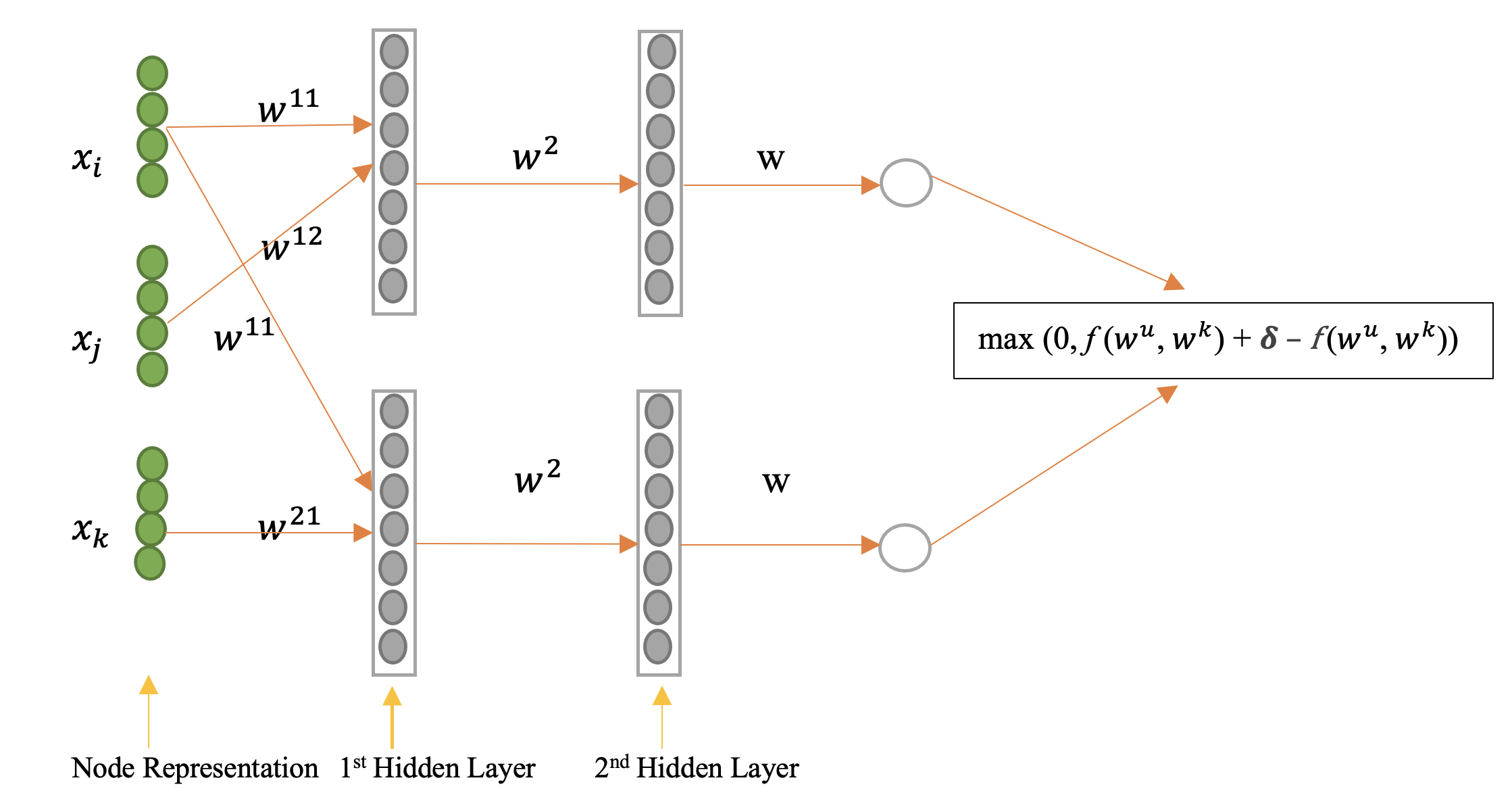}
\caption{ \raggedright The  architecture of SiNE \cite{wang2017signed} neural net. The network contains the embeddings of nodes. The SiNE trained on mini-batch stochastic gradient descent and update parameters backward direction. The backpropagation optimizes the deep network by propagating errors backward to calculate gradient.}
\end{figure}

Signed Graph Convolution Networks (SGCN)\cite{derr2018signed} also used extended balance theory. SGCN used positive and negative links based on balanced theory in an aggregation process. The theory classifies cycles in a signed network as being either balanced or unbalanced, whereas balanced theory (or unbalanced) cycle consists of an even (or odd) number of negative links. Similarly, a balanced path (or unbalanced path) consists of an even (or odd) number of negative links. According to balance theory, even number of negative links between two nodes are replaced with a positive link along a path of length l.  

\begin{figure}[tbh]
\includegraphics[width=8cm]{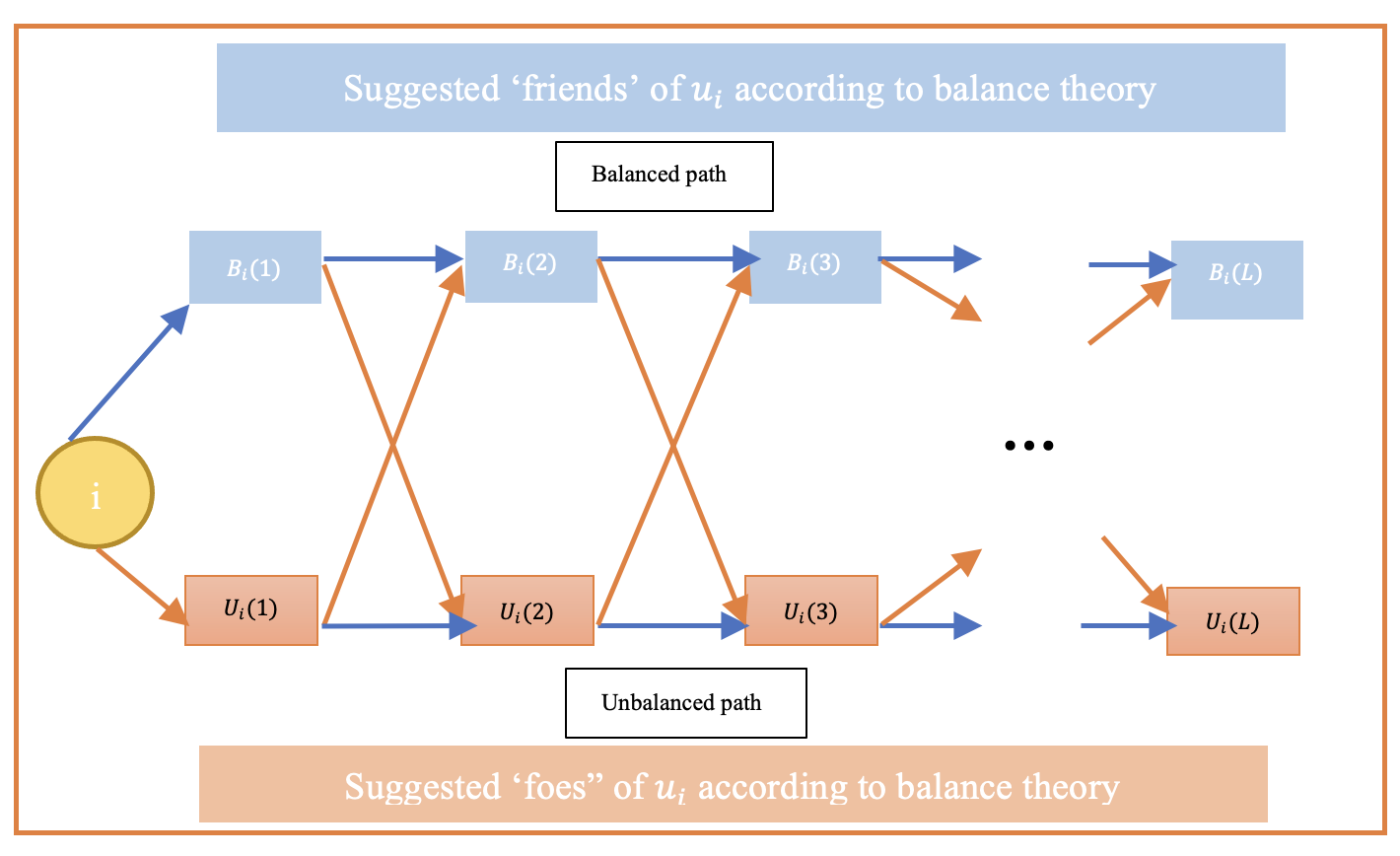}
\caption{\raggedright Illustration of signed network link structure \cite{derr2018signed}, the blue nodes are friends nodes and blue arrows define friendship. Red nodes are foes and red arrows define "foe" relationship. This structure is used in different aggregations based on balanced and unbalanced paths. }
\end{figure}

In Figure 10, we can see that from node $i$ to some other node along path $l$, considered a balanced path having all the positive link neighbors. From node $i$ to another node along path length $l$, adding a negative link to a balanced path, an unbalanced path is obtained here, \\
\text{where} \hspace{0.3cm} $l = 1$,\\
$B_i(1) = \{u_i|u_j \in \mathbb{N}_i^+ \}; U_i(1) = \{u_i|u_j \in \mathbb{N}_i^- \}$ \\  \\
$\text{For l} \hspace{0.3cm} \ge 1$, \\

\begin{equation}
\begin{aligned}
B_i(1+1) =  & \{u_j|u_k \in B_i(l) \hspace{0.3cm} \text{and} \hspace{0.3cm} u_j \in \mathbb{N}_k^+ \} \\
            & \cup \{u_j|u_k \in U_i^(l) \hspace{0.3cm} \text{and} \hspace{0.3cm} u_j \in \mathbb{N}_k^- \}  \\
U_i(1+1) =  & \{u_j|u_k \in U_i(l) \hspace{0.3cm} \text{and} \hspace{0.3cm} u_j \in \mathbb{N}_k^+ \} \\
            & \cup \{u_j|u_k \in B_i^(l) \hspace{0.3cm} \text{and} \hspace{0.3cm} u_j \in \mathbb{N}_k^- \}      
\end{aligned}
\end{equation}

GCNs aggregate and propagate information in signed networks using balanced and unbalanced sets/paths. SGCN keeps separate representations of nodes "friends" and "foes" because a single representation will not give a thorough idea.

\begin{figure}[tbh]
\includegraphics[width=8cm]{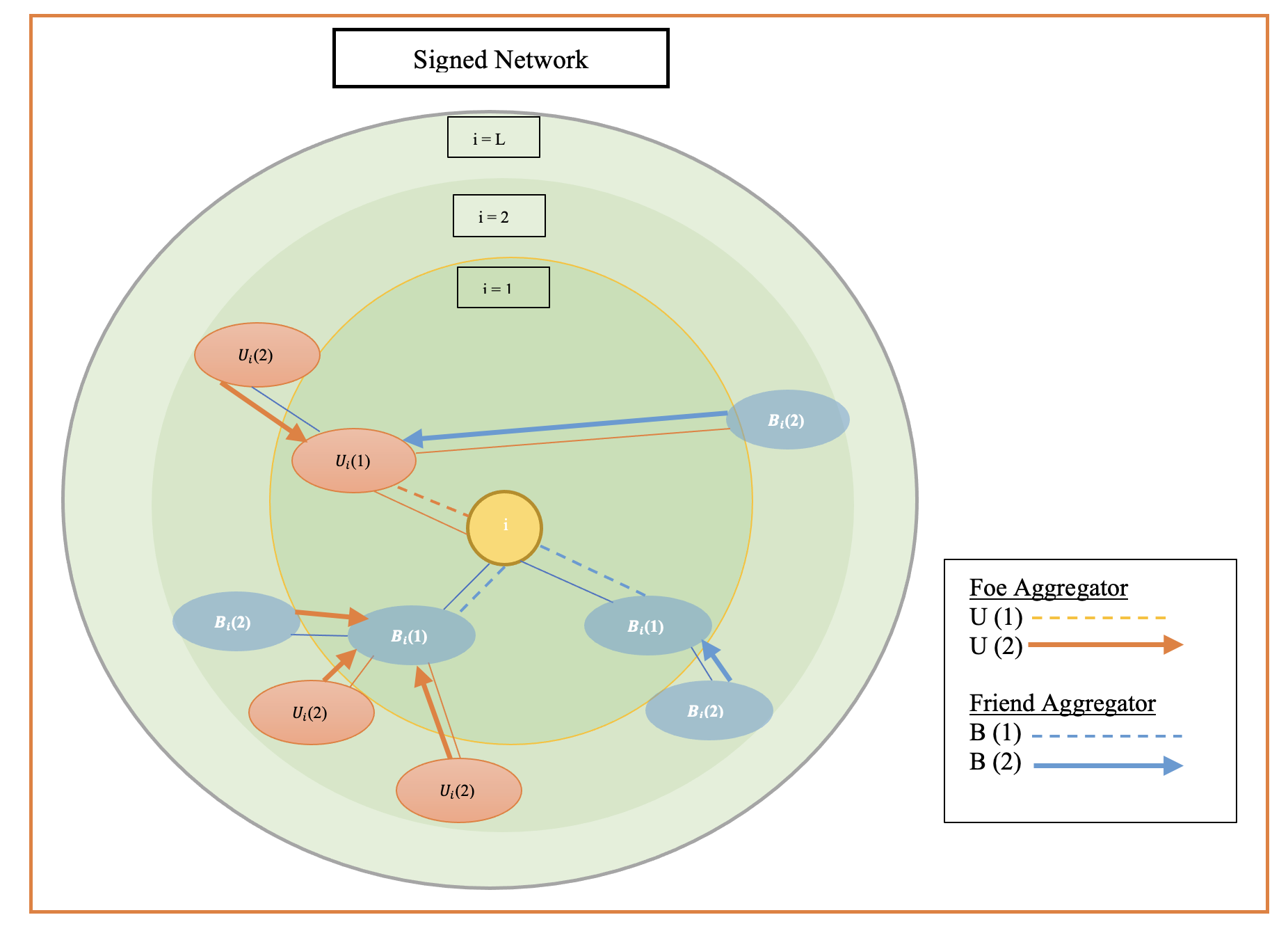}
\caption{ \raggedright Illustration of signed network link structure \cite{derr2018signed}, the blue nodes are friends nodes and blue arrows define friendship red nodes are foes and red arrows define "foes" relationship. This structure is used in different aggregations based on balanced and unbalanced paths.}
\end{figure}

In Figure 11, each cycle is labeled as level = 1,2,..., L. The aggregation and propagation process happens from level to level towards the $u_i$ node. For example, $u_i$ incorporates friend representation using $B(1)$ aggregator from the first layer.   
it is important to do the aggregation process correctly to stick to the balance theory. Therefore a second set of aggregators namely $B(2)$ and $U(2)$ propagates the information from users in sets $B_i(2)$ and $U_i(2)$, respectively. For example, here aggregator $U(2)$ seeks to utilize the information from users along the path by aggregating information from one positive and one negative link which is an unbalanced cycle. 
Here, level numbers l = 1,2,...,L denotes the number of cycles away the user from $u_i$ and simultaneously denotes at which layer in the signed GCN the user’s information will be incorporated. Separate aggregators are used here for both balanced and unbalanced sets.


The below defines the aggregation for $l \ge 1$:        

\begin{equation}  \label{eu_eqn}
\begin{aligned}
h_i^{B(l)} = \sigma (W^{B(l)} [ \sum_{j \in N_{i} ^ {+}} \frac{h_{j}^{B(l-1)}}{|N_{i}^{+}|}, \sum_{k \in N_{i} ^ {-}} \frac{h_{k}^{U(l-1)}}{|N_{i}^{-}|}, h_i^{B(l-1)} ]) \\
h_i^{U(l)} = \sigma (W^{U(l)} [ \sum_{j \in N_{i} ^ {+}} \frac{h_{j}^{U(l-1)}}{|N_{i}^{+}|}, \sum_{k \in N_{i} ^ {-}} \frac{h_{k}^{B(l-1)}}{|N_{i}^{-}|}, h_i^{U(l-1)} ])
\end{aligned}
\end{equation}

Here, $W^B(l)$, $W^U(l) \in \mathbb{R}^{d^{out}x2d^{in}}$ are for "friends" and "foes". The same logic is applicable to friends and foes representation. Here. at layer l, $h_i^{B(l)}$ represents all the aggregated positively linked neighbors information and $h_i^{U(l)}$ represents negatively collected linked neighbors information.The aggregation of any (friend/foe) representation at current layer depends on previous layer for the same representation. For example, friend representation does not only depend on direct friend connection from current layers but also depends on friends of friends from previous layers. The objective function of SGCN is based on two terms, the first term performs a multinomial linear regression (MLG) classifier to classify relative between pair of nodes and the second term performs extended balance theory to keep positively linked users closer in the embedding space. It also keeps no linked nodes closer than negatively linked nodes.

\subsubsection{\textbf{\large Graph Neural Network}} 
Neural networks have been successfully applied to grid-like structures, however, in the case of graph-like structures like biological nets, social nets, and telecommunications nets cannot be represented as grid-like structures. Graph Neural Network (GNN) is a powerful deep learning representation to deal with such kind of structure. GNN produces output for each node based on an iterative way. In neural nets, attention mechanisms have become a de facto standard as they allow dealing with variable-sized inputs and focus on the most relevant parts of the input to make decisions. Self-attention uses a single sequence to compute representation. GAT \cite{velivckovic2017graph} proposed an attention mechanism that is a single-layer feedforward neural network. To stabilize the learning process, GAT showed the extended version of self-attention which employs multi-head attention (K = 3) is beneficial. The K-independent attention mechanism concatenates their features to generate output feature representation. 

\begin{figure}[!tbh]
\includegraphics[width= 9cm]{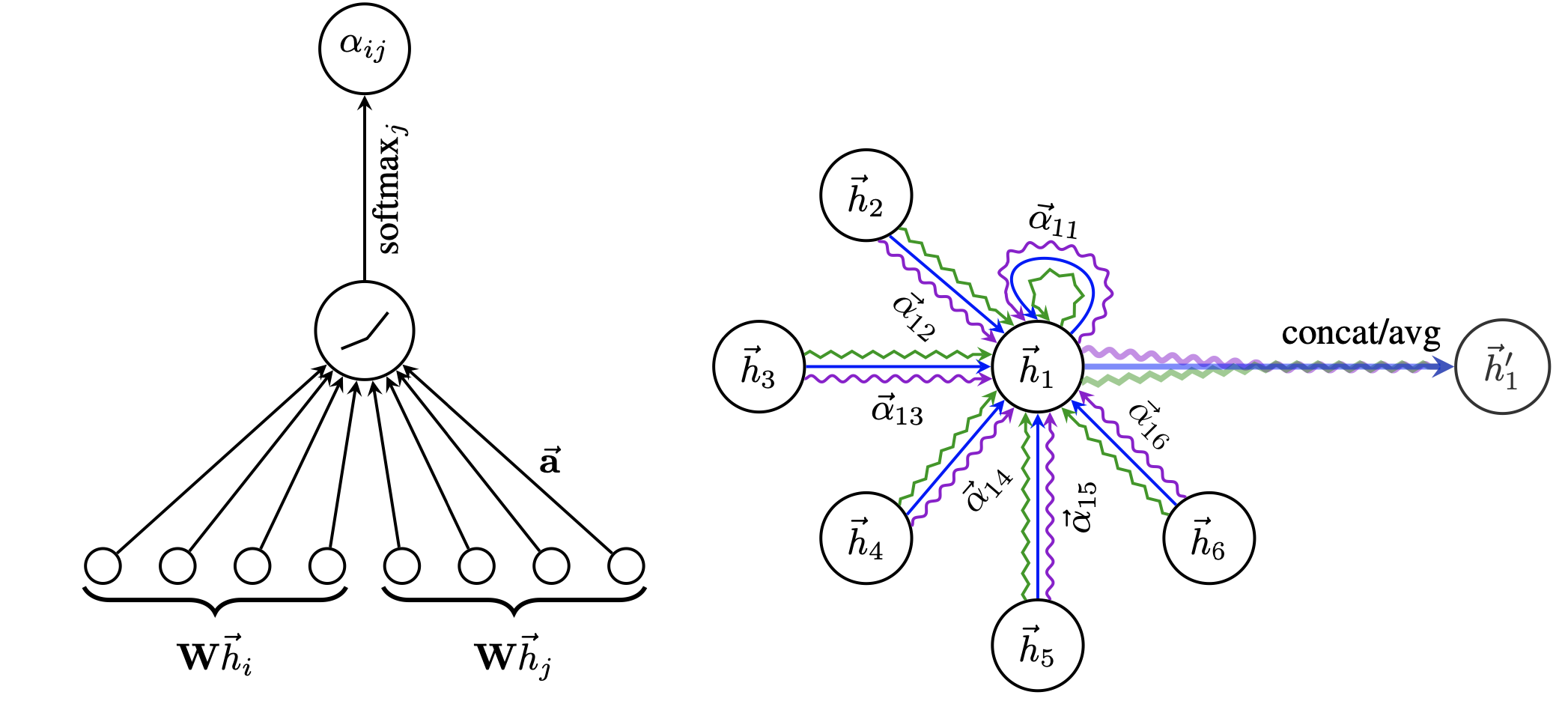}
\caption{\raggedright The figure adapted from \cite{velivckovic2017graph}, the left image illustrates the attention mechanism with a LeakyReLU activation. The right image illustrates multi-head attention (with k = 3 heads) by node 1 in its neighborhoods. The different arrow styles and colors denote independent attention computation. The aggregated features from each head are concatenated to obtain the output features.}
\end{figure}

At an advanced level, SiGAT \cite{huang2019signed} brought Graph Attention Network in signed graph embedding by capturing both balance theory and status theory. SiGAT outperforms SGCN on link prediction tasks. 

In a directed network, direction contains important information for status theory. So, the triads have different effects that are described by different motifs. SiGAT defined 2 motifs for undirected positive/negative neighbors, 4 motifs for positive/negative direction neighbors, and 32 different triangle motifs. The motifs are extracted from the graph at first and then applied to SiGAT model. 

GAT\cite{velivckovic2017graph} introduced an attention mechanism that uses a weight matrix to characterize different effects of different nodes on the target node. GAT computes $\alpha_{ij}$ for node i and j using attention mechanism \textbf{a} and LeakyReLU nonlinearity as: 

\begin{equation}  \label{eu_eqn}
\alpha_{ij} = \frac{exp(LeakyReLU(\textbf{a}^T[{Wh_i}||{Wh_j}))}{\sum_{k \in N_i} exp (LeakyReLU (\textbf{a}^T[{Wh_i}||{Wh_k}]))}
\end{equation}

here, $.^T$ and $||$ represents transposition and concatenation operation correspondingly. $\mathbb{N}$ is the set of neighbors of $i, \textbf{W} $ is the weight matrix parameter and $h_i$ is the node feature of node i.  

\begin{figure}[tbh]
\includegraphics[width= 9cm]{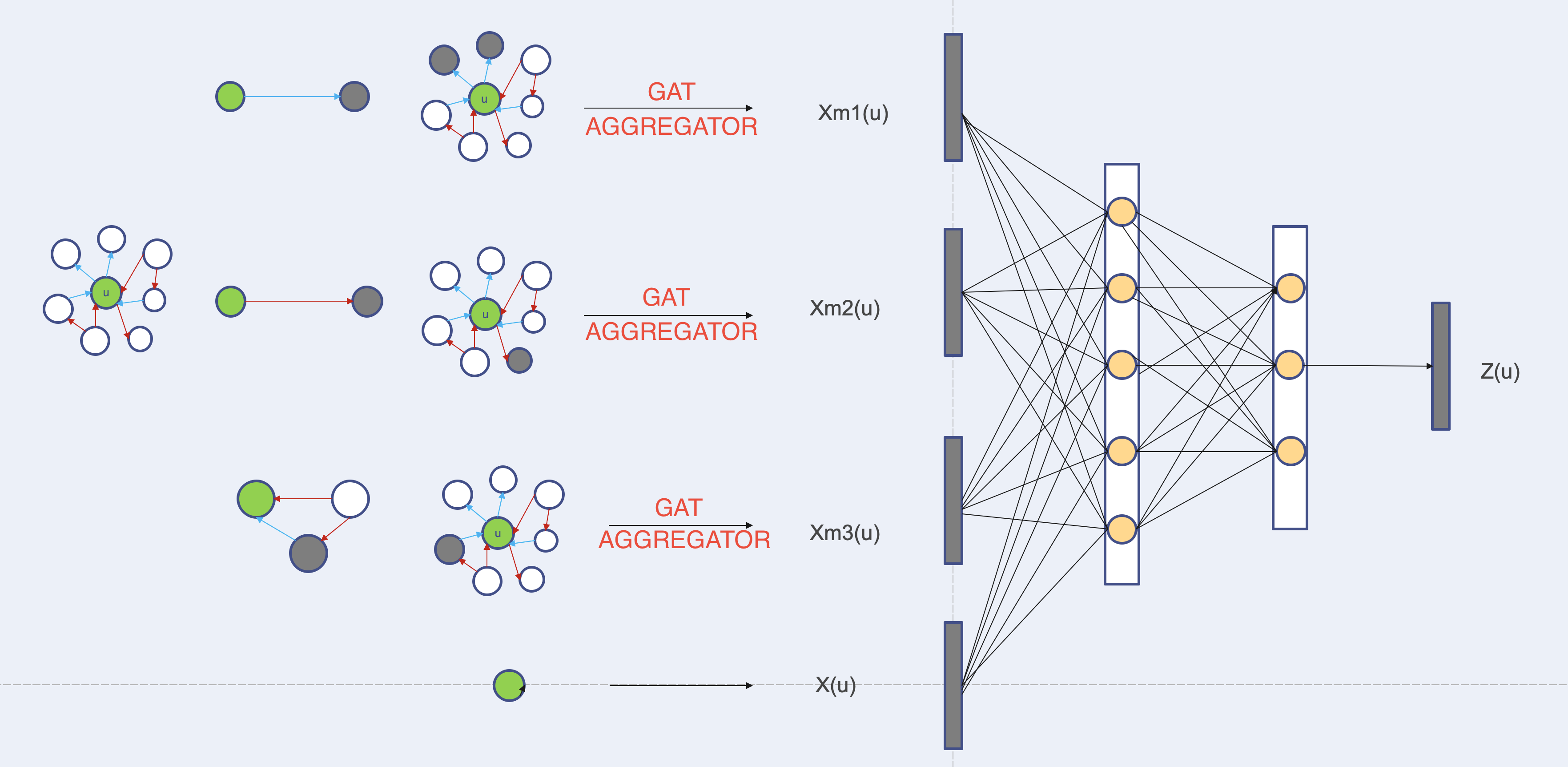}
\caption{$\raggedright$ Illustration of SiGAT \cite{huang2019signed} architecture, the blue arrows define friendship and red arrows define "foes" relationship. The gray color nodes are neighbors of node u under motif definition. After aggregating different motif information, concatenating $X_{mi}(u)$ and $X(u)$, the $Z(u)$ representation is obtained.}
\end{figure}

The concept is used in SiGAT to describe the influence of neighborhoods of a node u using different motifs. SiGAT gives different neighborhoods for different motifs and a GAT-AGGREGATOR aggregates all the information. A two-layer fully connected neural network concatenates all the messages from different motifs $X_{m1},X_{m2},X_{m3}, ...$ of a node u and obtain final node embedding $Z_u$. The loss is propagated backward to update parameters.

\begin{equation}  \label{eu_eqn}
\alpha_{uv}^{m_i} = \frac{exp(LeakyReLU(\textbf{a}^T_{m_i}[W_{m_i} X(u)||W_{m_i} X(v)))}{\sum_{k \in N_i} exp (LeakyReLU (\textbf{a}^T_{m_i}[W_{m_i} X(u)||W_{m_i} X(k)]))}
\end{equation}

\begin{equation}  \label{eu_eqn}
X_{m_i}(u) = \sum_{v \in N_{m_i}(u)} \alpha_{uv}^{mi}W_{m_i}X(v)
\end{equation}
(Equation 28) shows the final output feature of node u, after normalizing attention coefficients for node u and node v.  

SNEA \cite{li2020learning} utilized a masked self-attention layer which leverages the self-attention mechanism to aggregate neighbor information based on balance theory for node embedding. SNEA outperforms SIGAT in link prediction. Similar to SiGAT, here also a node is represented as balanced and unbalanced embeddings.

\begin{figure}[tbh]
\includegraphics[width= 9cm]{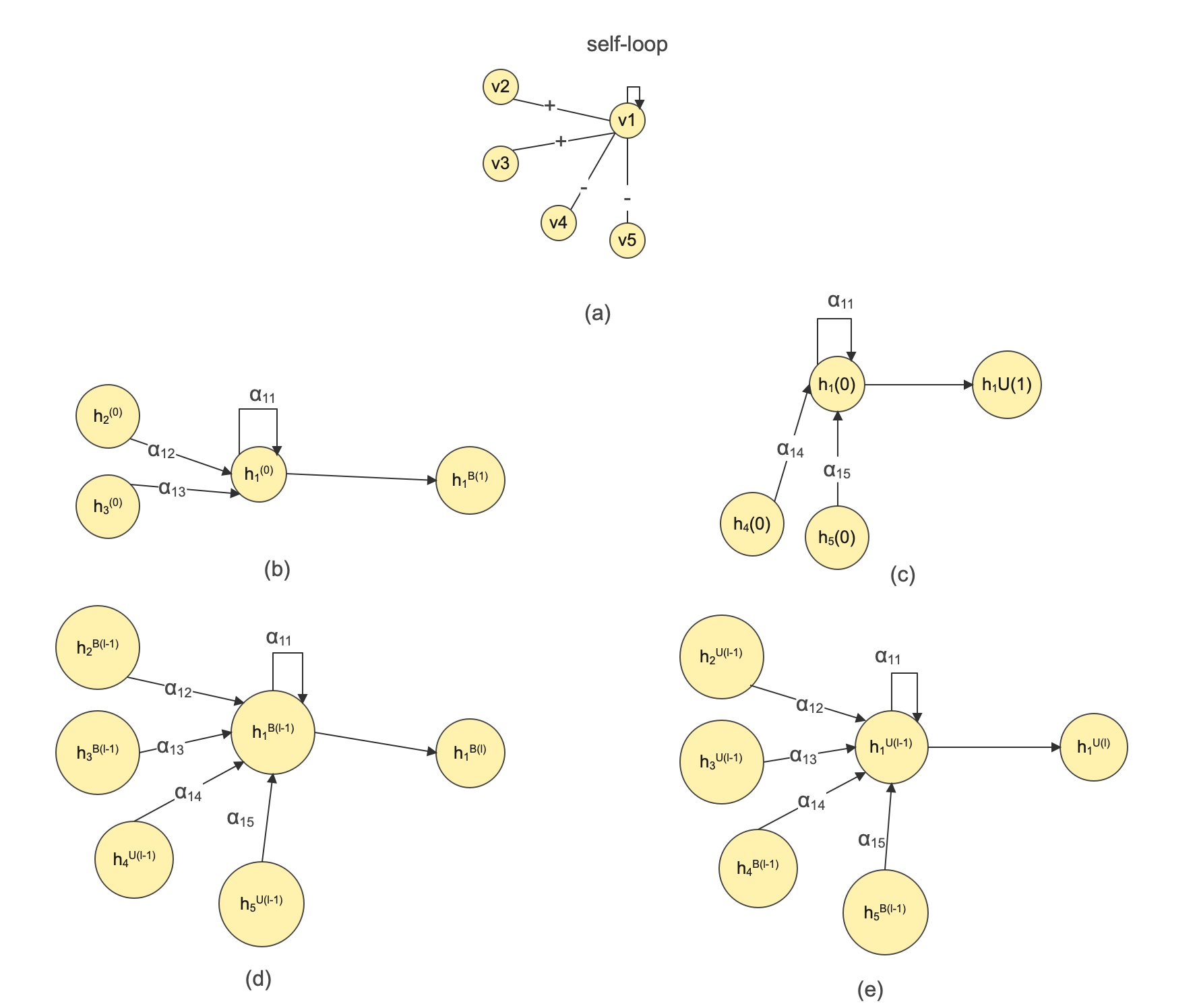}
\caption{\raggedright Illustration of SNEA \cite{li2020learning} architecture of self-attention mechanism and aggregation process from neighboring nodes. (a) A subnetwork (b)-(c) aggregation process for the first layer (d)-(e)aggregation process for deeper layers (i.e. l$\ge$1) }
\end{figure}

In SNEA, the attention layer is a single-layer feedforward neural net that is parameterized by a shared attentional parameter vector and a tanh function is applied to make the attention model non-linearity. For l $\ge$ 1 layers, deeper aggregation layers are recursively defined as:  

\begin{equation}  \label{eu_eqn}
\begin{aligned}
h_i^{B(l)} = tanh ( \sum_{j \in N_i^+,k \in N_i^- } \alpha_{ij}^{B(l)} h_j^{B(l-1)} W^{B(l)} +  \alpha_{ik}^{B(l)} h_k^{U(l-1)} W^{B(l)})
\end{aligned}
\end{equation}

\begin{equation}  \label{eu_eqn}
h_i^{U(l)} = tanh ( \sum_{j \in N_i^+,k \in N_i^- } \alpha_{ij}^{U(l)} h_j^{U(l-1)} W^{U(1)} + \alpha_{ik}^{U(l)} h_k^{B(l-1)} W^{U(1)})
\end{equation}

The aggregation process for $h_i^{B(l)}$ and $h_i^{U(l)}$ are the same. For the balanced embeddings $h_i^{B(l)}$, it aggregates balanced embeddings from balanced node set and unbalanced embeddings from unbalanced node set; and for unbalanced embeddings $h_i^{U(l)}$, aggregation process combines unbalanced embeddings from balanced set and balanced embeddings from unbalanced set. Here, $\alpha_{ij}^{B(l)}$ and $\alpha_{ij}^{U(l)}$ are attention scores of nodes for balanced and unbalanced sets and $W^{B(i)}, W^{U(i)}\in R^{_{in}^{(l)} X d_{out}^{(l)}}$ are shared linear transformation matrices.  
The normalized attention scores are calculated as follows:

\begin{equation}
\resizebox{5cm}{!}{
$\alpha_{ij}^{B(l)} = \frac{\exp(e_{ij}^{B(l)})}{\sum_{t \in N_i^+ \cup N_i^-} \exp(e_{ij}^{B(l)})}$
}
\label{eu_eqn}
\end{equation}
\begin{equation}  
\resizebox{4cm}{!}{
$\alpha_{ij}^{U(l)} = \frac{\exp(e_{ij}^{U(l)})}{\sum_{t \in N_i^+ \cup N_i^-} \exp(e_{ij}^{U(l)})}$
}
\label{eu_eqn}
\end{equation}


\textit{Summary:} In early work, convolution neural networks focused on employing the entire network. The SGCN designed the learning representation on the node level.  As we can that the incremental update on capturing  balance theory and status theory improved performance on experimental tasks.

\begin{table*}[h!] \setlength\tabcolsep{0.1pt} 
\caption{ Typical Signed graph embedding methods}
\begin{adjustbox}{width={\textwidth}}
\centering
\renewcommand{\arraystretch}{2}
\resizebox{19cm}{!}{\begin{tabular}{||l|l|l|l|l||}
\hline
\textbf{Method} & \textbf{Information} & \textbf{Task} & \textbf{Technique} & \textbf{Characteristic}  \\
 \hline\hline
SGCN\cite{derr2018signed} & Link Structure & Embedding & Convolution & Aggregation and Propagation\\
 \hline
SiNE\cite{wang2017signed} & Structural Balance Theory & Embedding & Deep Learning & Objective Function\\
\hline
SNE\cite{yuan2017sne} & Similarity Measure & Embedding & Log-bilinear & Signed network\\
\hline
SIGNet\cite{islam2018signet} & Structural Balance Theory & 2-dimensional embedding & Random walk & Weighted Signed directed and undirected \\
 \hline
SNEA\cite{wang2017attributed} & Extended Structural Balance Theory  & Embedding  & Optimization framework & Signed link and attributed user\\
 \hline
SiHet\cite{rizi2020signed} & Proximity between vertex & Embedding & Joint probability of different networks & Signed Network, Social Network and Signed Heterogeneous Network\\
\hline
DNE-SBP \cite{shen2018deep} & Structural link balance property & Embedding & Semi-supervised stacked auto-encoder  & Signed Network\\
\hline
SIDE\cite{kim2018side} &  Balance theory & Embedding & Random walk and proximity measure  & Signed directed Network\\
\hline
SHINE \cite{wang2018shine} & Structural balance theory & Embedding & Similarity measurement function and auto-encoder  & Signed sentiment network, social network and profile network\\
\hline
ComPath \cite{dhelim2020compath} & Proximity between vertex & Embedding & Similarity measurement function using proximity function & Heterogeneous Signed network\\
\hline
SGNS \cite{wang2018shine} & Structural balance theory and similarity measure & Embedding & Generative adversarial net & Signed undirected network\\
\hline
SNEGAN \cite{wang2018shine} & Structural balance theory and similarity measure &  Embedding & Signed random walk and similarity score & Signed network\\
\hline
StEM \cite{rahaman2018method} & Distance based ranking &  Embedding & Neural network & Signed network\\
\hline
nSNE \cite{song2018learning} & Distance based ranking &  Embedding & Node proximity and neural network & Unsigned and signed undirected network\\
\hline
CSNE \cite{mara2020csne} & Structural balance &  Embedding & Maximum Entropy &  Signed undirected network\\
\hline
SiGAT \cite{huang2019signed} & Balance and status theory &  Embedding & Attention Network &  Signed directed network\\
\hline

\end{tabular}}
\end{adjustbox}
\end{table*}

\section{\LARGE Experimental Tasks}
What are the research questions researchers ask to analyze signed graphs? Signed graph embedding brought a new dimension to analyze various downstream tasks similar to unsigned graph. Signed graph analysis also includes node-level, and edge-level predictions. Based on the data mining tasks, we categorize and summarize here four main data mining tasks.   
\subsection{\large Sign Link Prediction}
Link prediction is the most fundamental downstream task that has been thoroughly analyzed in graph embedding for signed graphs. Link prediction is even more important in social networks where links change; for example, links may form or dissolute on Twitter by following or unfollowing users in the future. In signed graph, link prediction predicts new positive, negative, or no edge between nodes.  In supervised methods, link prediction is a classification task. Most of the signed graph embedding models predict links based on a classification model. For example, SNE \cite{yuan2017sne} introduced a logistic regression model that can make use of node representations to compose edge representations to build the classifier. SNE compared four different types of element-wise operators for combining node vectors with edge vectors. SNE observed Hadamard operator achieves the highest accuracy over average, L1-Weight and L2-Weight for balanced dataset. Similarly, StEM  compared concatenation, Hadamard, L1, L2, and average to predict signs on missing or unobserved edges. 

Sign prediction has been analyzed in two ways, (i) predicts sign of existing links and (ii) predicts sign of unobserved links between two nodes. Most of the signed link prediction tasks in signed graph have focused on the first one. To predict signs of unobserved links between nodes can serve ground truth as existing links of original network are known. In supervised manner, a binary classification model is trained using labeled edges from balanced or unbalanced training data. SHINE \cite{wang2018shine}, SiHet \cite{rizi2020signed} selects a balanced test set and uses the rest of the network for training purposes whereas GCN, SiNE use the unbalanced test set for evaluation. Since edges involve two nodes, to compose edge representation two learned node representations are concatenated. In \cite{hu2019sparse}, cosine similarity of the vector representations of two nodes is used. If the similarity ranges from -1 it means an enemy relationship, 1 means a friend relationship, and 0 indicates no relationship. 

\begin{equation} \label{eu_eqn}
Similarity(u,v) = \frac{f(u).f(v)}{|f(u)|.f(v)|}
\end{equation}

SiNE trained a logistic regression classifier on training dataset which predicts links on test data. For comparison SiNE referred couple of baseline methods: Spectral clustering algorithm \cite{kunegis2010spectral} for signed version of Laplacian matrix where top-d eigenvectors where chosen corresponding to the smallest eigenvalues as low dimensional vector representation of nodes, FE \cite{leskovec2010predicting} degree based feature extraction method from signed social network, Matrix factorization \cite{hsieh2012low} based link prediction from two low-rank latent matrices. SNEA \cite{li2020learning} derives link feature combines two embedding of connected nodes. SiGAT \cite{huang2019signed} concatenates two learned representations to compose edge representation and then trains logistic regression model to predict edge sign in test set. 





\section{\LARGE Benchmark Datasets and Source Code}
In previous works, most of the research on signed networks applied their proposed models on four real world directed signed network dataset:
\begin{itemize}
\item{Bitcoin-Appha \cite{kumar2016structure}:This is who-trusts-whom network of people who trade using Bitcoin on a platform called Bitcoin Alpha.}
\item{Bitcoin-OTC \cite{kumar2016structure}:This is who-trusts-whom network of people who trade using Bitcoin on a platform called Bitcoin OTC.}
\item{Slashdot \cite{leskovec2010signed}:Slashdot is a technology-related news website know for its specific user community. The website features user-submitted and editor-evaluated current primarily technology oriented news. }
\item{Epinions \cite{leskovec2010signed}:This is who-trust-whom online social network of a a general consumer review site Epinions.com.}

\end{itemize}

\begin{table*}[h!] \setlength\tabcolsep{0.1pt} 
\caption{A summary of commonly used signed graph datasets}
\centering
\scalebox{0.85}{
\def\checkmark{\tikz\fill[scale=0.4](0,.35) -- (.25,0) -- (1,.7) -- (.25,.15) -- cycle;} 
\renewcommand{\arraystretch}{2}

\begin{tabular}{|p{2.7cm}|p{2.5cm}|p{2.5cm}|p{3cm}|p{2.5cm}|p{6cm}|}
\hline

\textbf{Dataset} & \textbf{Link Prediction} & \textbf{Node Clustering} & \textbf{Community Detection} & \textbf{Node Classification} & \textbf{Related Papers} \\
 \hline\hline
 
Bitcoin-Alpha \cite{kumar2016structure} & $\checkmark$ & $\checkmark$ &   &  &\cite{li2020learning,mara2020csne,grover2016node2vec,rahaman2018method,derr2018signed} \\ \hline
Bitcoin-OTC \cite{kumar2016structure} & $\checkmark$ &  & $\checkmark$ &   &\cite{li2020learning,mara2020csne,derr2018signed,rahaman2018method} \\ \hline
Slashdot \cite{leskovec2010signed} & $\checkmark$ &  &  & $\checkmark$&  \cite{dhelim2020compath,wang2017attributed,li2020learning,kim2018side,wang2017signed,yuan2017sne,mara2020csne,song2018learning,grover2016node2vec,rahaman2018method,kim2018side,derr2018signed}  \\ \hline
Epinions \cite{leskovec2010signed} & $\checkmark$  & $\checkmark$  & $\checkmark$ & $\checkmark$   &\cite{wang2017attributed},\cite{dhelim2020compath,li2020learning,kim2018side,wang2017signed,kim2018side,mara2020csne,grover2016node2vec,song2018learning,rahaman2018method,derr2018signed} \\ \hline
Wiki \cite{leskovec2010signed} & $\checkmark$  &  &  &$\checkmark$  &\cite{kim2018side,yuan2017sne,mara2020csne,song2018learning,wang2018shine,rahaman2018method} \\ \hline
Newsfullness \cite{leskovec2010signed} &  &  & $\checkmark$ &  &\cite{dhelim2020compath} \\ \hline
\end{tabular}}
\end{table*}
\indent
In our work, we are using citation network for different kind of research domains. Currently, we have explored Data Mining conferences and Software Engineering conferences. The dataset have been extracted from Google Scholar from timeline 2012 to 2022. We have collected 3000 research articles from each domain.  We have created authorship network from the dataset and currently building the citation network based on the papers have been cited in the 3000 research articles. 

\begin{table*}[h!] \setlength\tabcolsep{0.1pt} 
\caption{Source code of related models}
\centering
\scalebox{0.75}{
\renewcommand{\arraystretch}{2}

\begin{tabular}{|p{2.5cm}|p{15cm}|p{2.5cm}|}
\hline

\textbf{Paper} & \textbf{Source Code} & \textbf{Programming Platform} \\
 \hline\hline
 SGCN\cite{derr2018signed} & \url{https://github.com/benedekrozemberczki/SGCN}  & PyTorch \\
 \hline
\hline
\hline
SIGNet\cite{islam2018signet} & \url{https://github.com/raihan2108/signet} & Python and C++ \\
 \hline
 \hline
SiHet\cite{rizi2020signed} & \url{https://github. com/fatemehsrz/SiHet} & Python\\
\hline
DNE-SBP \cite{shen2018deep} & \href{https://github.com/shenxiaocam/Deep-network-embedding-for-graph-representation-learning-in-signed-networks}{DNE-SBP} & MATLAB \\
\hline
SIDE\cite{kim2018side} &  \url{https://datalab.snu.ac.kr/side/} & Python \\
\hline
StEM \cite{rahaman2018method} & \url{https://github.com/InzamamRahaman/StEMPublic} &  Python \\
\hline
nSNE \cite{song2018learning} & \url{https://github.com/wzsong17/Signed-Network-Embedding} &  PyTorch \\
\hline
SiGAT \cite{huang2019signed} & \url{https://github.com/huangjunjie95/SiGAT} &  Python \\
\hline
\end{tabular}}
\end{table*}

\section{\LARGE CHALLENGES AND FUTURE WORKS}
Graph learning has got huge attention to embed graphs in low-dimensional space for graph analysis program. In future we want to analyze heterogeneous signed graph using graph embedding technique to explore downstream tasks. In this section, we discuss the challenges and possible future research directions. 

\subsection{Create Signed Graph network}
As a direction, a detailed study of how citations are made within each research domain by differentiating citations into endorsement(positive), criticism(negative), and neutral categories. That is a way to create signed network between papers and authors from a social network perspective. In each paper, there are hundreds of citations listed in the reference. The paper citations can be categorized into these three types. For the classification task, we will follow the keyword-based technique to label the citation sentiment as positive, negative, or neutral from a list of positive and negative words \cite{wilson2005recognizing}, which was used for opinion finding system. It is important to create the network in such a way that follows balance and status theory to do further analysis from social network perspective. 

\subsection{Capturing Signed Graph Properties for Graph Embedding}
In future work, we aim to explore more advanced features of authorship and citation networks. Specifically, we plan to extract meaningful features from both the citation network and the author network, with a focus on learning the underlying structures of these networks. Since graph embedding methods primarily emphasize structural aspects, understanding these dynamics will be crucial.

Furthermore, we propose the construction of a heterogeneous signed citation network, where authors and papers are represented as two distinct types of nodes, and the links between them reflect positive and negative interactions. In addition to the structural properties of the network, we hypothesize that the network’s features—such as its temporal evolution—can provide deeper insights into its dynamics. For example, observing how a real-world citation graph changes over time can reveal patterns of influence and collaboration within the scientific community.

A key direction for future research will be embedding such dynamic citation networks within a deep learning framework. By integrating the temporal evolution of citation graphs, we believe there is significant potential to improve the performance and accuracy of signed citation graph embeddings. This approach could open new avenues for understanding the behavior of citation networks and enhancing predictive models for graph-based tasks.

\section{\LARGE CONCLUSIONS}
Signed graph embedding has been significantly explored for graph analysis in recent times. Incorporating signed citation network will give the research community new insights. This survey conducts a comprehensive study of the state-of-the-art signed graph embedding methods. In this work, we have summarized the methods and applications of signed graphs and discussed the datasets and resources in a systematic way. We have also explored some applications as our future work plan. We hope this survey provides a clean sketch on signed graph embedding which can help researchers pursue their research interest in this field. 

\bibliographystyle{plain} 
\bibliography{ref.bib} 

\addtolength{\textheight}{-12cm}   




\section*{ACKNOWLEDGMENT}

I would like to thanks my advisor, Dr. Bojan Cukic, Professor of Computer Science Department  University of North Carolina at Charlotte for his constant encouragement towards the realization of this work.

\end{document}